\documentclass[epj]{svjour}
\usepackage{graphics}

\begin{document}
\title{Kinetic Exchange Models for Income and Wealth Distributions}

\author{
Arnab Chatterjee \inst{1}
\thanks{\emph{Present Address:} Condensed Matter and
Statistical Physics Section,
The Abdus Salam International Centre for Theoretical Physics,
Strada Costiera 11, Trieste I-34014, Italy.
\emph{Email:} \texttt{achatter@ictp.it}
}
\and
Bikas K. Chakrabarti \inst{1,2}
\thanks{\emph{Email:}
\texttt{bikask.chakrabarti@saha.ac.in}}
}
\institute{Theoretical Condensed Matter Physics Division and
Centre for Applied Mathematics and Computational Science,\\
Saha Institute of Nuclear Physics, 1/AF Bidhannagar, Kolkata 700064, India.
\and Economic Research Unit,
Indian Statistical Institute, 203 B. T. Road, Kolkata 700108, India.
}

\abstract{Increasingly, a huge amount of statistics have been gathered which
clearly indicates that income and wealth
distributions in various countries or societies follow a robust pattern,
close to the Gibbs distribution of energy in an ideal gas in equilibrium.
However, it also deviates in the low income and more 
significantly for the high income ranges.
Application of physics models provides illuminating ideas
and understanding, complementing the observations.}


\PACS{89.20.Hh,89.75.Hc,89.75.Da,43.38.Si}

\maketitle
\section{Introduction}\label{sec:1}
In any society or country, if one can isolate its count on people and their
money or wealth, one finds that while the total money or wealth 
remains fairly constant on a relatively longer time scale, its movement
from individual to individual is not so
due to its dynamics at shorter time scales (daily or weekly).
Eventually, on overall average for the society or
country, there appears very robust money or wealth distributions. 
Empirical data for society show a small variation in the value of the 
power-law exponent that characterises the `tail' of the distribution, 
while it equals to unity for firms.
Locally, of course, there appear many `obvious
reasons' for such uneven distribution of wealth or income
within the societies.
However, such `reasons' seem to be very ineffective if the global
robust structure of the income and wealth distribution in various
societies is considered. Statistical physics based models for such
distributions seem to succeed in capturing the essential `reasons'
for such universal aspects of the distributions.

Here, we review the empirical basis for considering
kinetic exchange models for income and wealth distributions
in Sec.~\ref{sec:empirical}. We then discuss the gas like models
in Sec.~\ref{sec:idealgas} and give the details of their numerical
analyses in Sec.~\ref{sec:ds:numerical}, while in Sec.~\ref{sec:anaystudy}
we review the analytical studies done so far on these models.
Sec.~\ref{sec:othmodel} discusses other model studies,
including an annealed savings model and a model with a
non-consumable commodity. 
Finally, we end with discussions in Sec.~\ref{sec:disc}.

\section{Empirical studies of income and wealth distributions}
\label{sec:empirical}
\begin{figure*}
\resizebox{0.99\columnwidth}{!}{
\includegraphics{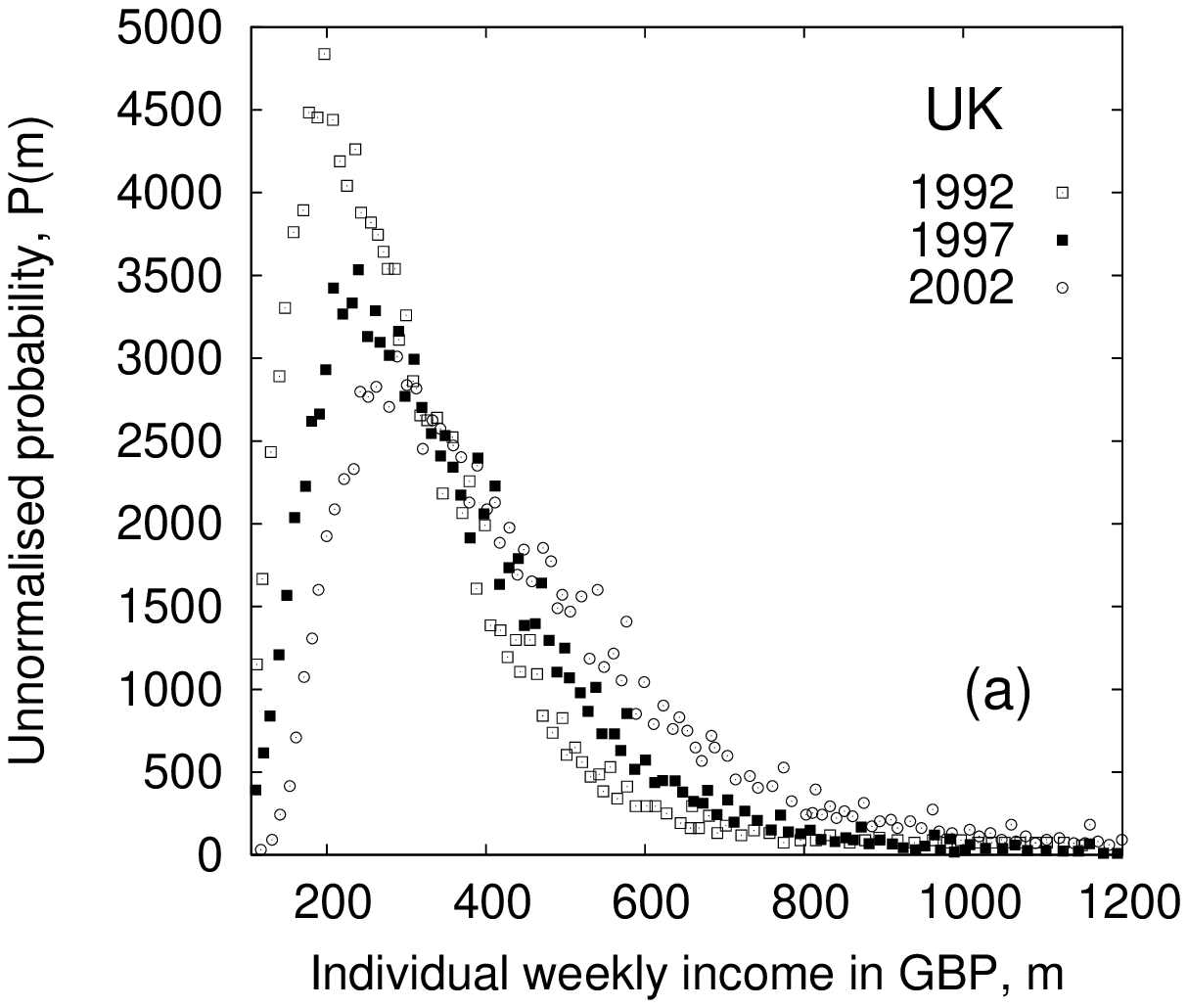}
}
\resizebox{0.99\columnwidth}{!}{
\includegraphics{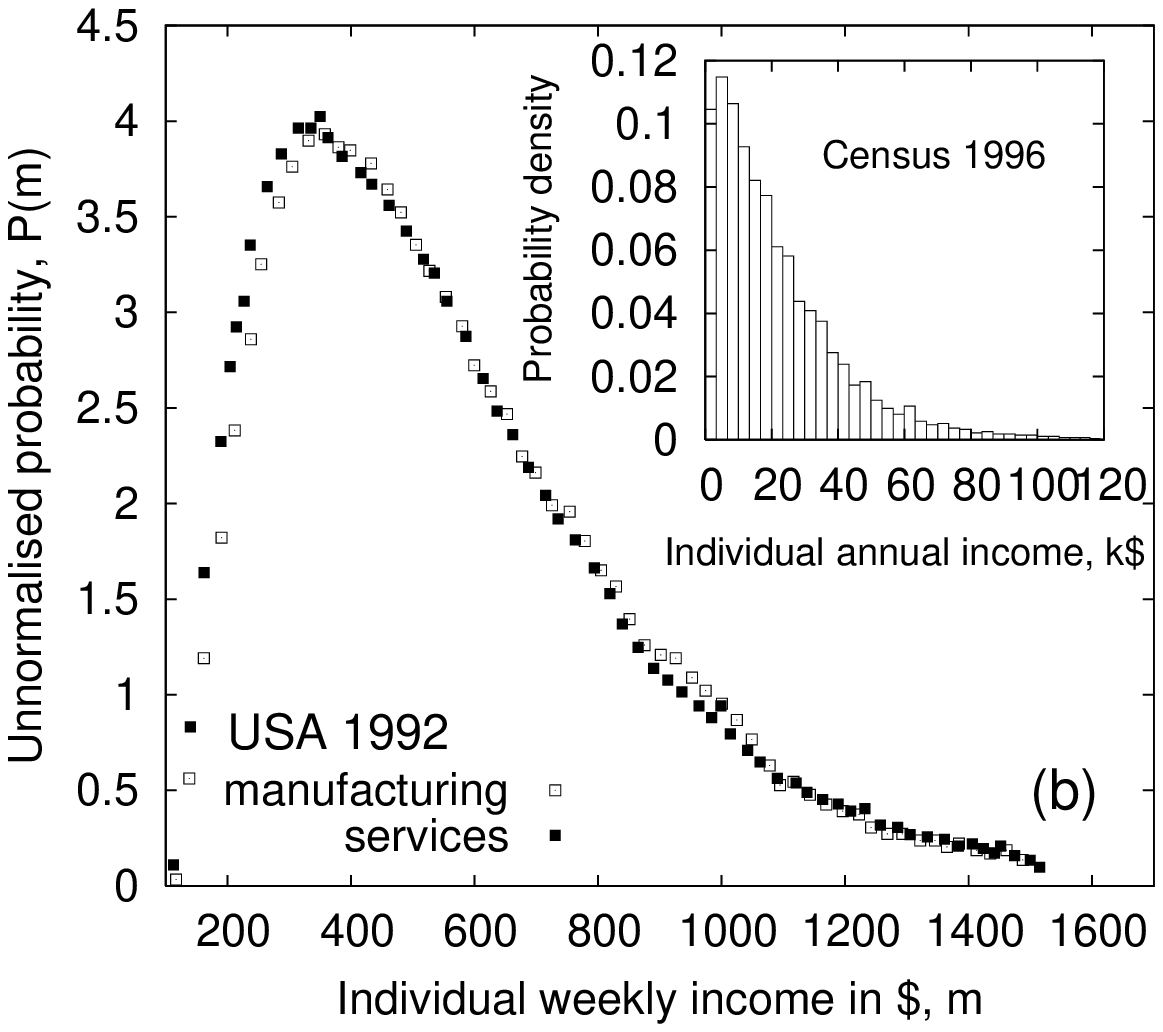}
}
\resizebox{0.99\columnwidth}{!}{
\includegraphics{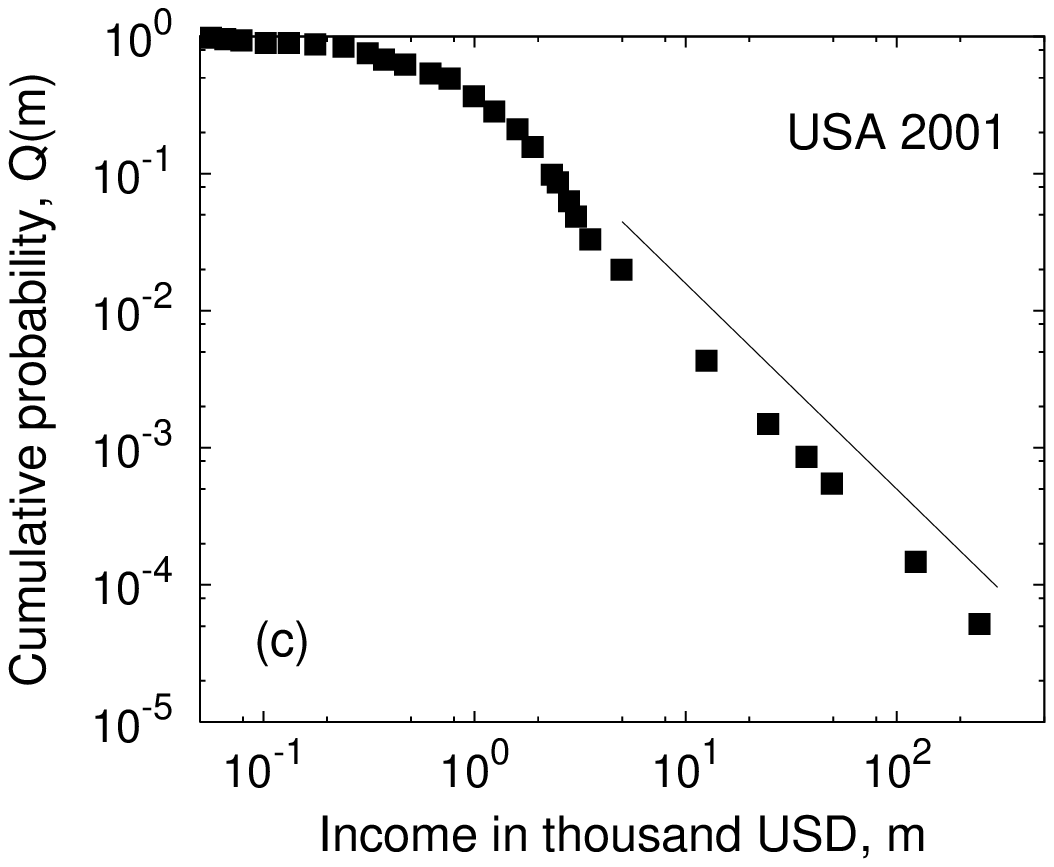}
}
\resizebox{0.99\columnwidth}{!}{
\includegraphics{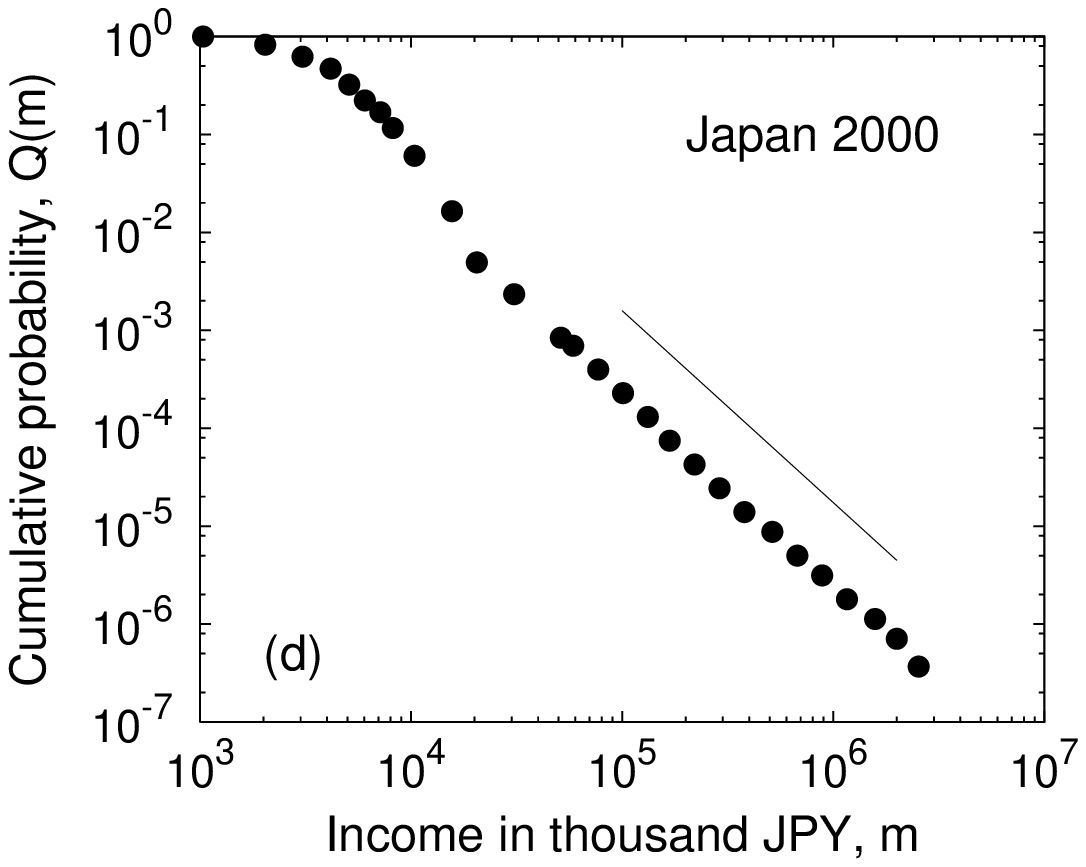}
}
\caption{
(a) Distribution $P(m)$ of individual weekly income in UK for
1992, 1997 and 2002; data adapted from Ref.~\cite{Willis:2004}. 
(b) Distribution $P(m)$ of individual weekly income for
manufacturing and service sectors in USA for 1992; 
data for US Statistical survey, taken from Ref.~\cite{Willis:2004}. 
The inset shows the probability distribution of individual annual 
income, from US census data of 1996. The data is adapted from 
Ref.~\cite{DraguYakov01a}.
(c) Cumulative probability $Q(m) = \int_m^\infty P(m) dm$ 
of rescaled adjusted gross personal annual income in US for IRS data from 2001 
(adapted from Ref~\cite{Yakodata}),
with Pareto exponent $\nu \approx 1.5$
(given by the slope of the solid line).
(d) Cumulative probability distribution of Japanese personal income
in the year 2000 (data adapted from Ref.~\cite{Fujidata}). 
The power law (Pareto) region approximately fits to $\nu=1.96$.
}\label{fig:realdataset}
\end{figure*}

The distribution of wealth among individuals in an economy has been an
important area of research in economics, for more than a hundred
years~\cite{Pareto:1897,Mandelbrot:1960,EWD05,ESTP}. 
The same is true for income distribution in any society.
Detailed analysis of the income distribution~\cite{EWD05,ESTP} so far
indicate
\begin{equation}
\label{par}
P(m) \sim 
\left\{ \begin{array}{lc}
m^\alpha \exp(-m/T) & \textrm{for} \ m < m_c,\\
m^{-(1+\nu)} \ \ \ \ & \textrm{for} \  m \ge m_c,
\end{array}\right.
\end{equation}
where $P$ denotes the number density of people with income 
or wealth $m$ and $\alpha$, $\nu$ denote exponents and $T$ denotes a scaling
factor. The power law in income and wealth distribution (for $m \ge m_c$) is
named after Pareto and the exponent $\nu$ is called the Pareto exponent.
A historical account of Pareto's data and that from recent sources can be
found in Ref.~\cite{Richmond:ESTP}.
The crossover point ($m_c$) is extracted from the crossover from the 
Gamma distribution form to the power law tail. One often fits the
region below $m_c$ to a log-normal form 
$\log P(m) \propto -(\log m)^2$.
Although this form is often preferred by economists, we think that the other
Gamma distribution form Eqn.~(\ref{par}) fits better with the data,
because of the remarkable fit with the Gibbs distribution in 
Ref.~\cite{Yakodata}.
We consider that in the following discussion.
This robust feature of $P(m)$ seems to be very well established for the
analysis of the enormous amount of data available 
today (See Fig.~\ref{fig:realdataset}).
We consider this distribution (in view of its stability and universality)
to be an `equilibrium' (in the thermodynamic sense) distribution in a
many-body (interacting, statistical) system like a gas, where the Gibbs
distribution are established for more than 100 years. This paper
reviews the various attempts in this direction.

Although Pareto~\cite{Pareto:1897} and Gini~\cite{Gini} had respectively 
identified the power-law tail and the log-normal bulk of the income 
distribution, the demonstration of both features in the same
distribution was possibly first demonstrated by Montroll and 
Shlesinger~\cite{MontrollShlesinger} through an analysis
of fine-scale income data obtained from the US Internal Revenue Service 
(IRS) for the year 1935-36. It was observed that while the top 2-3 \% 
of the population (in terms of income) followed a power law with Pareto 
exponent $\nu \simeq 1.63$; the rest followed a lognormal
distribution. Later work on Japanese personal income data based on 
detailed records obtained from the Japanese National Tax Administration 
indicated that the tail of the distribution followed a power law  value that 
fluctuated from year to year around the mean value of 2~\cite{Aoyama:2000}.
Further work~\cite{Souma:2000} showed that the power law region described the 
top 10\% or less of the population (in terms of income), while the remaining 
income distribution was well-described by the log-normal form. While the 
value fluctuated significantly from year to year, it was observed that the 
parameter describing the log-normal bulk, the Gibrat 
index~\cite{Gibrat:1931}, remained 
relatively unchanged. The change of income from year to year, i.e., the
growth rate as measured by the log ratio of the income tax paid in 
successive years, was observed by Fujiwara et al~\cite{Fujiwara:2003} 
to be also a heavy tailed distribution, although skewed, and
centered about zero. Later work on the US income distribution based 
on data from IRS for the years 1997-1998, while still indicating a 
power-law tail (with $\nu \simeq 1.7$), have suggested
that the the lower 95\% of the population have income whose distribution 
may be better described by an exponential 
form~\cite{DraguYakov01a,Dragulescu:2001}. 
The same observation has been made for income distribution in the UK 
for the years 1994-1999, where the value  was found to vary
between $2.0$ and $2.3$, but the bulk seemed to be well-described by 
an exponential decay.

It is interesting to note that, when one shifts attention from the
income of individuals to the income of companies, one still observes 
the power law tail. A study of the income distribution of Japanese firms
\cite{Okuyamaetal} concluded that it follows a power law with $\nu \simeq 1$,
which is also often referred to as the Zipf's law. Similar
observation has been reported for the income distribution of US companies
\cite{Axtell}.

Compared to the empirical work done on income distribution, relatively few
studies have looked at the distribution of wealth, which consist of the net
value of assets (financial holdings and/or tangible items) owned at
a given point in time.
The lack of an easily available data source for measuring wealth,
analogous to income tax returns for measuring income, means that one
has to resort to indirect methods. Levy and Solomon~\cite{Levy:1997} used
a published list of wealthiest people to generate a rank-order distribution,
from which they inferred the Pareto exponent for wealth distribution
in USA.
Refs.~\cite{Dragulescu:2001} and
\cite{Coelho:2004} used an alternative technique based on adjusted data
reported for the purpose of inheritance tax to obtain the Pareto
exponent for UK.
Another study used tangible asset (namely house area) as a measure of
wealth to obtain the wealth distribution exponent in ancient Egyptian
society during the reign of Akhenaten (14th century BC)~\cite{Egypt}.
The wealth distribution in Hungarian medieval society has
also been seen to follow a Pareto law~\cite{Hegyi:2007}.
More recently, the wealth distribution in India at present was also
observed to follow a power law tail with the exponent varying 
around $0.9$~\cite{Sinha:2006}.
The general feature observed in the limited empirical study of
wealth distribution is that of a power law behavior for the wealthiest
$5-10 \%$ of the population, and exponential or log-normal
distribution for the rest of the population.
The Pareto exponent as measured from the wealth distribution is found
to be always lower than the exponent for the income distribution,
which is consistent with the general observation that, in market
economies, wealth is much more unequally distributed than
income~\cite{Samuelson:1980}.

The striking regularities (see Fig.~\ref{fig:realdataset}) 
observed in the income distribution for different countries, have 
led to several new attempts at 
explaining them on theoretical grounds. Much of the current impetus is
from physicists' modelling of economic behavior in analogy with
large systems of interacting particles, as treated, e.g., in the kinetic
theory of gases. According to physicists working on this problem, the
regular patterns observed in the income (and wealth) distribution may
be indicative of a natural law for the statistical properties of a
many-body dynamical system representing the entire set of economic 
interactions in a society, analogous to those previously derived for
gases and liquids. By viewing the economy as a thermodynamic system,
one can identify the income distribution with the distribution of
energy among the particles in a gas.
In particular, a class of kinetic exchange models have provided a simple
mechanism for understanding the unequal accumulation of assets.
Many of these models, while simple from the perspective of economics,
has the benefit of coming to grips with the key factor in socioeconomic
interactions that results in very different societies converging to
similar forms of unequal distribution of resources (see 
Refs.~\cite{EWD05,ESTP}, which consists of a collection of 
large number of technical papers in this field; see 
also~\cite{Hayes:2002,Hogan,Ball,Carbone:2007,Gallegati:2006,PhysOrg} 
for some popular discussions and criticisms as well as appreciations).

Considerable investigations with real data during the last ten years revealed
that the tail of the income distribution indeed follows the above
mentioned behavior and the value of the Pareto exponent $\nu$ is generally
seen to vary between 1 and 
3~\cite{Dragulescu:2001,Levy:1997,Sinha:2006,Oliveira:1999,Aoyama:2003,DiMatteo:2004,Clementi:2005,Clementi:EWD:2005,Ding:2007}. It is also
known that typically less than $10 \%$ of the population in any country
possesses about $40 \%$ of the total wealth of that country and they follow
the above law. The rest of the low income population, 
follow a different distribution which is debated 
to be either 
Gibbs~\cite{Dragulescu:2001,Levy:1997,Aoyama:2003,marjit,Ispolatov:1998,Dragulescu:2000} 
or log-normal~\cite{DiMatteo:2004,Clementi:2005,Clementi:EWD:2005}.

\section{Gas-like models}
\label{sec:idealgas}
In 1960, Mandelbrot wrote ``There is a great temptation to consider 
the exchanges of money which occur in economic interaction as 
analogous to the exchanges of energy which occur in physical shocks 
between molecules. In the loosest possible terms, both kinds 
of interactions {\it should} lead to {\it similar} states of 
equilibrium. That is, one {\it should} be able to explain the law 
of income distribution by a model similar to that used in 
statistical thermodynamics: many authors have done so explicitly, 
and all the others of whom we know have done so 
implicitly.''~\cite{Mandelbrot:1960}. However,  
Mandelbrot does not provide any references to this bulk of material!
Here, we discuss the recent literature and the developments.

In analogy to two-particle collisions with a resulting change 
in their individual kinetic energy (or momenta), income exchange 
models may be based on two-agent interactions. 
Here two randomly selected agents exchange 
money by some pre-defined mechanism. Assuming the exchange process 
does not depend on previous exchanges, the dynamics follows a 
Markovian process:
\begin{equation}
\left( \begin{array}{c}
m_i (t+1) \\
m_j (t+1)
\end{array}
\right) = {\mathcal M}
\left( \begin{array}{c}
m_i (t) \\
m_j (t)
\end{array}
\right)
\label{matrixM}
\end{equation}
where $m_i (t)$ is the income of agent $i$ at time $t$ and the 
collision matrix ${\mathcal M}$ defines the exchange mechanism.
\begin{figure}
\centering{
\resizebox*{8.0cm}{!}{\includegraphics{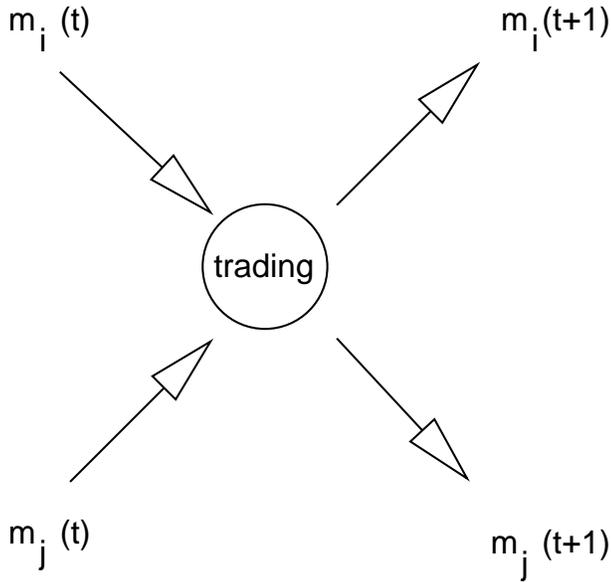}}
}
\caption{Schematic diagram of the trading process. Agents $i$ and $j$
redistribute their money in the market: $m_i(t)$ and $m_j(t)$, their
respective money before trading, changes over to $m_i(t+1)$ and $m_j(t+1)$
after trading.}
\label{fig1:scatter}
\end{figure}

In this class of models, one considers a closed economic system
where total money $M$ and total number of agents $N$ is fixed. 
This corresponds to a situation where no production
or migration occurs and the only economic activity is confined to trading.
Each agent $i$, individual or corporate, posesses money $m_i(t)$ at time $t$.
In any trading, a pair of traders $i$ and $j$ exchange their money
\cite{marjit,Ispolatov:1998,Dragulescu:2000,Chakraborti:2000}, 
such that their total money is (locally) conserved (Fig.~\ref{fig1:scatter})
and none end up with negative money ($m_i(t) \ge 0$, i.e, debt not allowed):
\begin{equation}
\label{mdelm}
m_i(t+1) = m_i(t) + \Delta m; \  m_j(t+1) = m_j(t) - \Delta m 
\end{equation}
following local conservation:
\begin{equation}
\label{consv}
m_i(t) + m_j(t) = m_i(t+1) + m_j(t+1);
\end{equation}
time ($t$) changes by one unit after each trading.

\subsection{Model A: Without any savings}
The simplest model considers a random fraction of total money
to be shared~\cite{Dragulescu:2000}:
\begin{equation}
\label{deltam}
\Delta m = \epsilon_{ij} [m_i(t) + m_j(t)] - m_i(t),
\end{equation}
where $\epsilon_{ij}$ is a random fraction ($\ 0 \le \epsilon_{ij} \le 1$) 
changing with time or trading.
The steady-state ($t \rightarrow \infty$) distribution of money is Gibbs one:
\begin{equation}
\label{gibbs}
P(m)=(1/T)\exp(-m/T);T=M/N.
\end{equation}
Hence, no matter how uniform or justified the initial distribution is, the
eventual steady state correspond to Gibbs a distribution where most of the
people have got very little money.
This follows from the conservation of money and additivity of entropy:
\begin{equation}
\label{prob}
P(m_1)P(m_2)=P(m_1+m_2).
\end{equation}
This steady state result is quite robust and realistic too!
In fact, several variations of the trading, and of the `lattice'
(on which the agents can be put and each agent trade with its
`lattice neighbors' only), whether compact, fractal or small-world like
\cite{Oliveira:1999}, leaves the distribution unchanged. Some other variations
like random sharing of an amount $2m_2$ only (not of $m_1 + m_2$)
when $m_1 > m_2$ (trading at the level of lower economic class in the trade),
lead even to a drastic situation: all the money in the market drifts to one
agent and the rest become truely pauper \cite{Chakraborti:2002,Hayes:2002}.

\subsection{Model B: With uniform savings}
\label{subsec:fixedsaving}
In any trading, savings come naturally \cite{Samuelson:1980}.
A saving propensity factor $\lambda$ was therefore introduced in the random 
exchange model~\cite{Chakraborti:2000} 
(see~\cite{Dragulescu:2000} for model without savings), where each trader
at time $t$ saves a fraction $\lambda$ of its money $m_i(t)$ and trades
randomly with the rest:
\begin{equation}
\label{fmi}
m_i(t+1)=\lambda m_i(t) + \epsilon_{ij} \left[(1-\lambda)(m_i(t) + m_j(t))\right],
\end{equation}
\begin{equation}
\label{fmj}
m_j(t+1)=\lambda m_j(t) + (1-\epsilon_{ij}) \left[(1-\lambda)(m_i(t) + m_j(t))\right],
\end{equation}
where
\begin{equation}
\label{eps}
\Delta m=(1-\lambda )[\epsilon_{ij} \{m_{i}(t)+m_{j}(t)\}-m_{i}(t)],
\end{equation}
$\epsilon_{ij}$ being a random fraction, coming from the stochastic nature
of the trading.
\begin{figure}
\resizebox{0.99\columnwidth}{!}{
\includegraphics{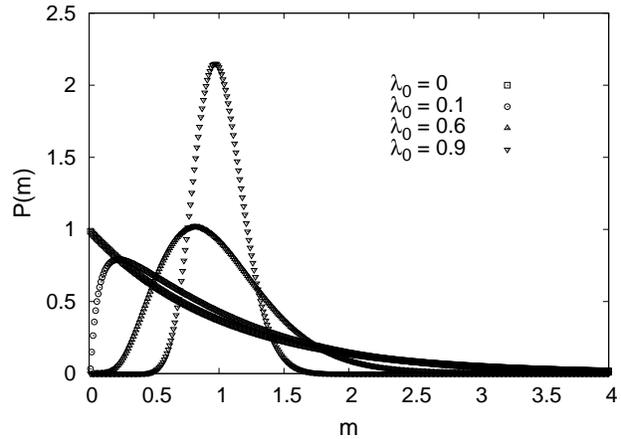}
}
\caption{Steady state money distribution $P(m)$ for the model
with uniform savings. The data shown are for different values
of $\lambda$: $0, 0.1, 0.6, 0.9$ for a system size $N=100$.
All data sets shown are for average money per agent $M/N = 1$.
}
\label{fig:fixedlam}       
\end{figure}

The market (non-interacting at $\lambda =0$ and $1$) becomes `interacting'
for any non-vanishing $\lambda (<1)$: For fixed $\lambda$ (same for all
agents), the steady state distribution $P(m)$ of money is exponentially
decaying on both sides with the most-probable money per agent shifting away
from $m=0$ (for $\lambda =0$) to $M/N$ as $\lambda \to 1$
(Fig.~\ref{fig:fixedlam}). This self-organizing feature of the market,
induced by sheer self-interest of saving by each agent without any global
perspective, is quite significant as the fraction of paupers decrease with
saving fraction $\lambda$ and most people end up with some finite 
fraction of the
average money in the market (for $\lambda \to 1$, the socialists'
dream is achieved with just people's self-interest of saving!).
Interestingly, self-organisation also occurs in such market models when there
is restriction in the commodity market~\cite{Chakraborti:2001}.
Although this fixed saving propensity does not give yet the Pareto-like
power-law distribution, the Markovian nature of the scattering or trading
processes (Eqn. (\ref{prob})) is effectively lost.
Indirectly through $\lambda$, the agents get to know (start interacting with)
each other and the system co-operatively self-organises towards a 
stable form with a non-vanishing 
most-probable income (see Fig.~\ref{fig:fixedlam}).

Patriarca et al~\cite{Patriarca:2004} 
claimed through heuristic arguments (based on
numerical results) that the distribution is a close approximate
form of the Gamma distribution
\begin{equation}
\label{gamma:pat}
P(m) = 
C
m^{\alpha} \exp[-m/T]
\end{equation}
where $T=1/(\alpha + 1)$ 
and $C=(\alpha +1)^{\alpha +1}/\Gamma(\alpha +1)$, $\Gamma$ 
being the Gamma function whose argument $\alpha$ is related
to the savings factor $\lambda$ as:
\begin{equation}
\label{gamma:lambda}
\alpha =   \frac{3 \lambda}{ 1- \lambda}.
\end{equation}
When compared with Eqn.~(\ref{gibbs}) for $\lambda=0$ limit, it is to be 
noted that $M/N = 1$ here. 
Also, when compared with Eqn.~(\ref{par}), $m_c \to \infty$.
the qualitative argument forwarded here~\cite{Patriarca:2004}
is that, as $\lambda$ increases, effectively the agents (particles)
retain more of its money (energy) in any trading (scattering).
This can be taken as implying that with increasing $\lambda$,
the effective dimensionality increases and temperature of the 
scattering process changes~\cite{Patriarca:2004}.

This result has also been supported by numerical results in Ref.~\cite{BMM}.
However, a later study~\cite{Repetowicz:2005,Richmond:2005} analyzed the moments, and
found that moments upto the third order agree with those obtained
from the form of the Eqn.~(\ref{gamma:lambda}), and discrepancies start
from fourth order onwards. Hence, the actual form of the distribution
for this model still remains to be found out.

It seems that a very similar model was proposed by 
Angle~\cite{Angle:1986,Lux:EWD:2005,Angle:2006} several years back in sociology journals.
Angle's `One Parameter Inequality Process' model 
is described by the equations:
\begin{eqnarray}
\label{eqn:angle}
m_i(t+1) &=& m_i(t) + D_t w m_j(t) - (1-D_t) w m_i(t) \nonumber\\
m_j(t+1) &=& m_j(t) + (1-D_t) w m_i(t) - D_t w m_j(t)\nonumber\\
\end{eqnarray}
where $w$ is a fixed fraction and $D_t$ takes value $0$ or $1$ randomly.
The numerical simulation results of Angle's model 
fit well to Gamma distributions.

In the gas like models with uniform savings, the distribution
of wealth shows a self organizing feature. A peaked distribution
with a most-probable value indicates an economic scale. Empirical
observations in homogeneous groups of individuals as in waged income
of factory labourers in UK and USA~\cite{Willis:2004}
and data from population survey in USA among students of different school
and colleges produce similar distributions~\cite{Angle:2006}. 
This is a simple case where a homogeneous population 
(say, characterised by a unique value of $\lambda$) has been identified.
\subsection{Model C: With distributed savings}
\label{subsec:mixedsaving}
In a real society or economy, 
the interest of saving varies from person to person, which implies that
$\lambda$ is a very inhomogeneous parameter.
To imitate this situation, we move a step closer to the real 
situation where saving factor $\lambda$ is
widely distributed within the population
\cite{Chatterjee:2004,Chatterjee:2003,Chakrabarti:2004}.
The evolution of money in such a trading can be written as:
\begin{equation}
\label{mi}
m_i(t+1)=\lambda_i m_i(t) + \epsilon_{ij} \left[(1-\lambda_i)m_i(t) + (1-\lambda_j)m_j(t)\right],
\end{equation}
\begin{equation}
\label{mj}
m_j(t+1)=\lambda_j m_j(t) + (1-\epsilon_{ij}) \left[(1-\lambda_i)m_i(t) + (1-\lambda_j)m_j(t)\right]
\end{equation}
The trading rules are same as before, except that
\begin{equation}
\label{lrand}
\Delta m=\epsilon_{ij}(1-\lambda_{j})m_{j}(t)-(1-\lambda _{i})(1 - \epsilon_{ij})m_{i}(t)
\end{equation}
here; where $\lambda_{i}$ and $\lambda_{j}$ are the saving
propensities of agents $i$ and $j$. The agents have fixed (over time) saving
propensities, distributed independently, randomly and uniformly (white)
within an interval $0$ to $1$: agent $i$ saves a random fraction
$\lambda_i$ ($0 \le \lambda_i < 1$) and this $\lambda_i$ value is quenched
for each agent ($\lambda_i$ are independent of trading or $t$).
Studies show that for uniformly distributed saving propensities,
$\rho(\lambda) = 1$ for $0 \le \lambda < 1$, one gets eventually
$P(m) \sim m^{(1+\nu)}$, with $\nu = 1$ (see Fig.~\ref{fig:distlambda}).
The eventual deviation from the power law in $Q(m)$ in the inset of
Fig.~\ref{fig:distlambda} is due to the exponential cutoff contributed
by the rare statistics for high $m$ value.
\section{Numerical analysis of models A, B and C}
\label{sec:ds:numerical}
\begin{figure}
\resizebox{0.99\columnwidth}{!}{
\includegraphics{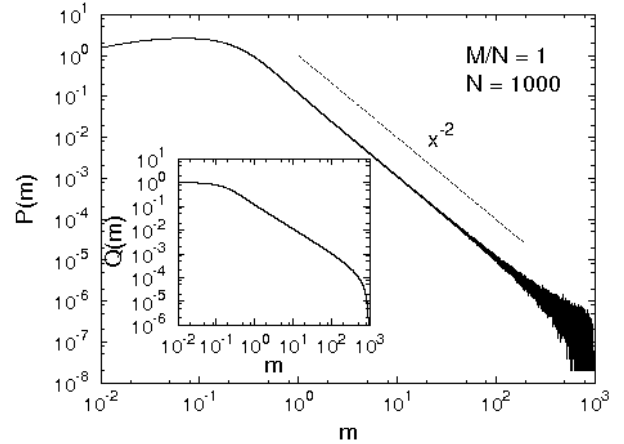}
}
\caption{Steady state money distribution $P(m)$ for the distributed
$\lambda$ model with $0 \le \lambda < 1$ for a system of $N=1000$ agents.
The $x^{-2}$ is a guide to the observed power-law, with $1+\nu=2$.
Here, the average money per agent $M/N = 1$.}
\label{fig:distlambda}       
\end{figure}
\begin{figure}
\resizebox{0.99\columnwidth}{!}{
\includegraphics{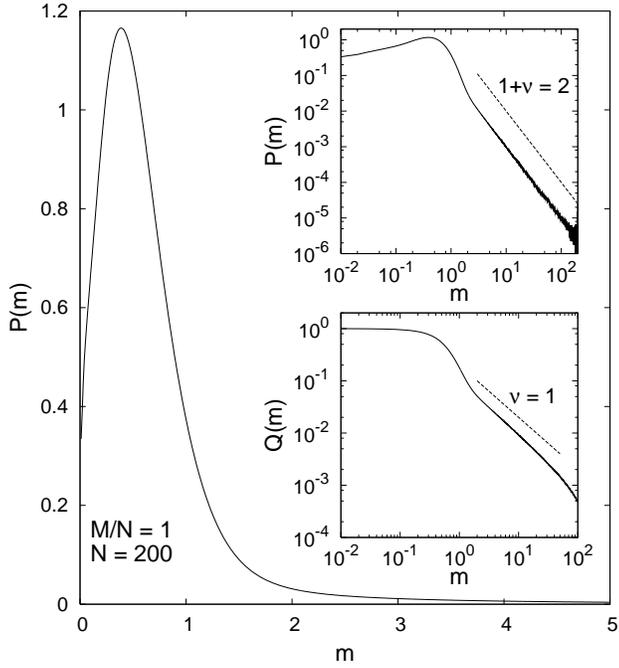}
}
\caption{Steady state money distribution $P(m)$ for a model with $f=0.6$
fraction of agents with a uniform saving propensity $\lambda_1 =0.6$
and the rest $1-f$ fraction having random uniformly distributed (quenched)
savings, in $0 \le \lambda < 1$ for a system of $N=200$ agents.
Here, the average money per agent $M/N = 1$.
The top inset shows $P(m)$ in log-log scale for the full range,
while the bottom inset shows the cumulative distribution $Q(m)$. 
In addition to the power law tail
in $P(m)$ and $Q(m)$ (as in the basic, distributed savings model),
$Q(m)$ resembles a behavior similar to observed in empirical 
data (see Fig.~\ref{fig:realdataset}).
}
\label{fig:mixlambda-fd}       
\end{figure}
Starting with an arbitrary initial (uniform or random) distribution of
money among the agents, the market evolves with the trading. At each time,
two agents are randomly selected and the money exchange among them occurs,
following the above mentioned scheme. We check for the steady state, by
looking at the stability of the money distribution in successive
Monte Carlo steps $t$ (we define one Monte Carlo time step as $N$ pairwise
exchanges). Eventually, after a typical relaxation time
the money distribution becomes stationary. 
This relaxation time is dependent on system size $N$ and the 
distribution of $\lambda$ 
(e.g, $\sim 10^6$ for $N=1000$ and uniformly distributed $\lambda$). 
After this, we average the money distribution over $\sim 10^3$
time steps. Finally we take configurational average over $\sim 10^5$
realizations of the $\lambda$ distribution to get the money distribution
$P(m)$. It is found to follow a 
power law for the wealthiest population ($\sim 10 \%$).
This decay fits to Pareto
law Eqn.~(\ref{par}) with $\nu \simeq 1$ (Fig.~\ref{fig:distlambda}). 
We also checked that for a mixed population where a fraction $f$
has fixed saving propensity $\lambda = \lambda_1$ and for the rest
($1-f$ fraction), $\lambda$ is distributed uniformly within
$0 \le \lambda < 1$, we find a money distribution resembling very much
the observed empirical distributions (see Fig.~\ref{fig:mixlambda-fd}),
as shown in Fig.~\ref{fig:realdataset}.
Here, when $P(m)$ is fitted in Eqn.~(\ref{par}), we have $\nu=1$
and the exponent $\alpha$ is approximately given by Eqn.~(\ref{gamma:lambda})
with $\lambda=\lambda_1$ and $m_c$ depending on $f$ and $\lambda_1$.
Note, for
finite size $N$ of the market, the distribution has a narrow initial growth
upto a most-probable value $m_p$ after which it falls off with a power-law
tail for several decades. This
Pareto law (with $\nu \simeq 1$) covers the entire range in $m$ of the
distribution $P(m)$ in the limit $N \rightarrow \infty$. We checked that
this power law is extremely robust: apart from the uniform $\lambda$
distribution used in the simulations in Fig.~\ref{fig:distlambda}, 
we also checked the results for a distribution
\begin{figure}
\resizebox{0.99\columnwidth}{!}{
\includegraphics{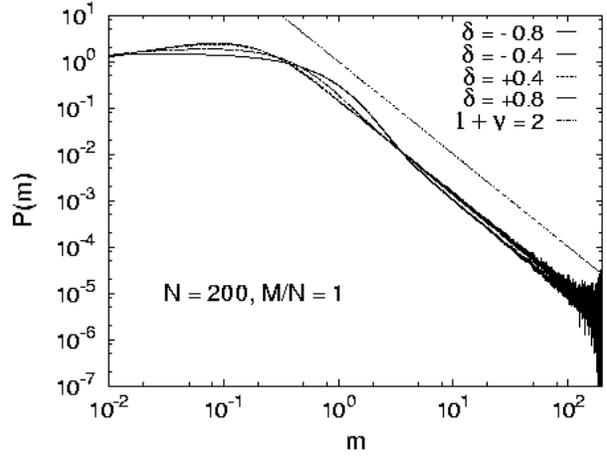}
}
\caption{Steady state money distribution $P(m)$ in the model for $N = 200$ 
agents with $\lambda$ distributed as  $\rho (\lambda) \sim \lambda^\delta$ 
with different values of $\delta$. 
A guide to the power law with exponent $1+\nu=2$ is also provided.
For all cases, the average money per agent $M/N = 1$.}
\label{fig:lamalpha}       
\end{figure}
\begin{figure}
\resizebox{0.99\columnwidth}{!}{
\includegraphics{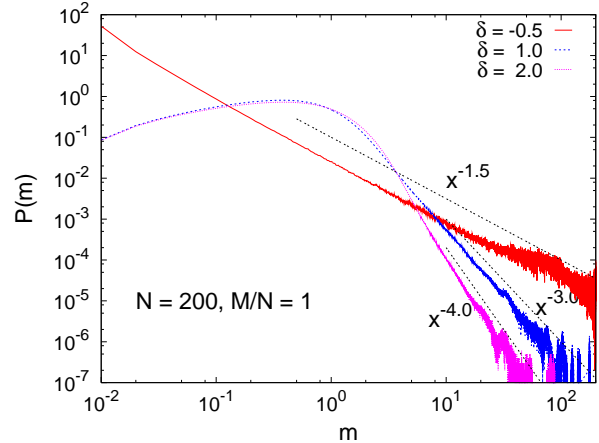}
}
\caption{Steady state money distribution $P(m)$ in the model for $N = 200$ 
agents with $\lambda$ distributed as  
$\rho (\lambda) \sim |1 - \lambda|^\delta$ 
with different values of $\delta$.
The distributions $P(m)$ have power law tails $P(m) \sim m^{-(1+\nu)}$,
where the power law exponents $1+\nu$ approximately equal to $2+\delta$
indicated by the dotted straight lines.
For all cases, the average money per agent $M/N = 1$.}
\label{fig:1-al}       
\end{figure}
\begin{equation}
\label{lam0}
\rho(\lambda) \sim |\lambda_0-\lambda|^\delta,\quad \lambda_0 \ne 1, \quad 0<\lambda<1,
\end{equation}
of quenched $\lambda$ values among the agents. The Pareto law with $\nu=1$ is
universal for all $\delta$. The data in Fig.~\ref{fig:distlambda} 
corresponds to
$\lambda_0 = 0$, $\delta = 0$. For negative $\delta$ values, however,
we get an initial (small $m$) Gibbs-like decay in $P(m)$ 
(see Fig.~\ref{fig:lamalpha}).
Fig.~\ref{fig:1-al} shows that for $\lambda_0 =1$, the resultant distribution
is $P(m) \sim m^{-(1+\nu)}, ~ \nu=1+\delta$.
\begin{figure}
\resizebox{0.99\columnwidth}{!}{
\includegraphics{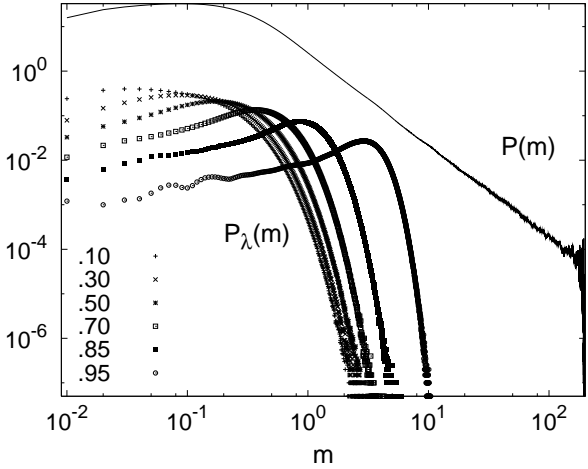}
}
\caption{Steady state money distribution $P_{\lambda}(m)$ 
for some typical values of $\lambda$ (= $0.1,0.3,0.5,0.7,0.85,0.95$)
in the distributed $\lambda$ model. The data is collected from the
ensembles with $N=200$ agents. The total distribution of money $P(m)$
is also plotted for comparison.
For all cases, the average money per agent $M/N = 1$.}
\label{fig:mlam}       
\end{figure}

In case of uniformly distributed saving propensity $\lambda$
($\rho(\lambda)=1$, $0 \le \lambda <1$), 
the individual money distribution $P_{\lambda}(m)$ for
agents with any particular $\lambda$ value, although differs considerably,
remains non-monotonic (see Fig.~\ref{fig:mlam}), 
similar to that for uniform $\lambda$ market
with $m_p(\lambda)$ shifting with $\lambda$ (see Fig.~\ref{fig:fixedlam}). 
Few subtle points may
be noted though: while for uniform $\lambda$ the $m_p(\lambda)$ were all
less than of the order of unity (average money per agent is fixed to
$M/N = 1$; see Fig.~\ref{fig:fixedlam}), 
for distributed $\lambda$ case $m_p(\lambda)$ can be considerably larger 
and can approach to the order of $N$ for large $\lambda$ 
(see Fig.~\ref{fig:mlam}). This in consistent with the empirically
known fact that the large-income people usually have larger
saving factors~\cite{Dynan:2004}.

There is also a marked qualitative difference in fluctuations:
while for fixed $\lambda$, the fluctuations in time (around the most-probable
value) in the individuals' money $m_i(t)$ gradually decreases with increasing
$\lambda$, for quenched distribution of $\lambda$, the trend gets reversed.
\begin{figure}
\resizebox{0.99\columnwidth}{!}{
\includegraphics{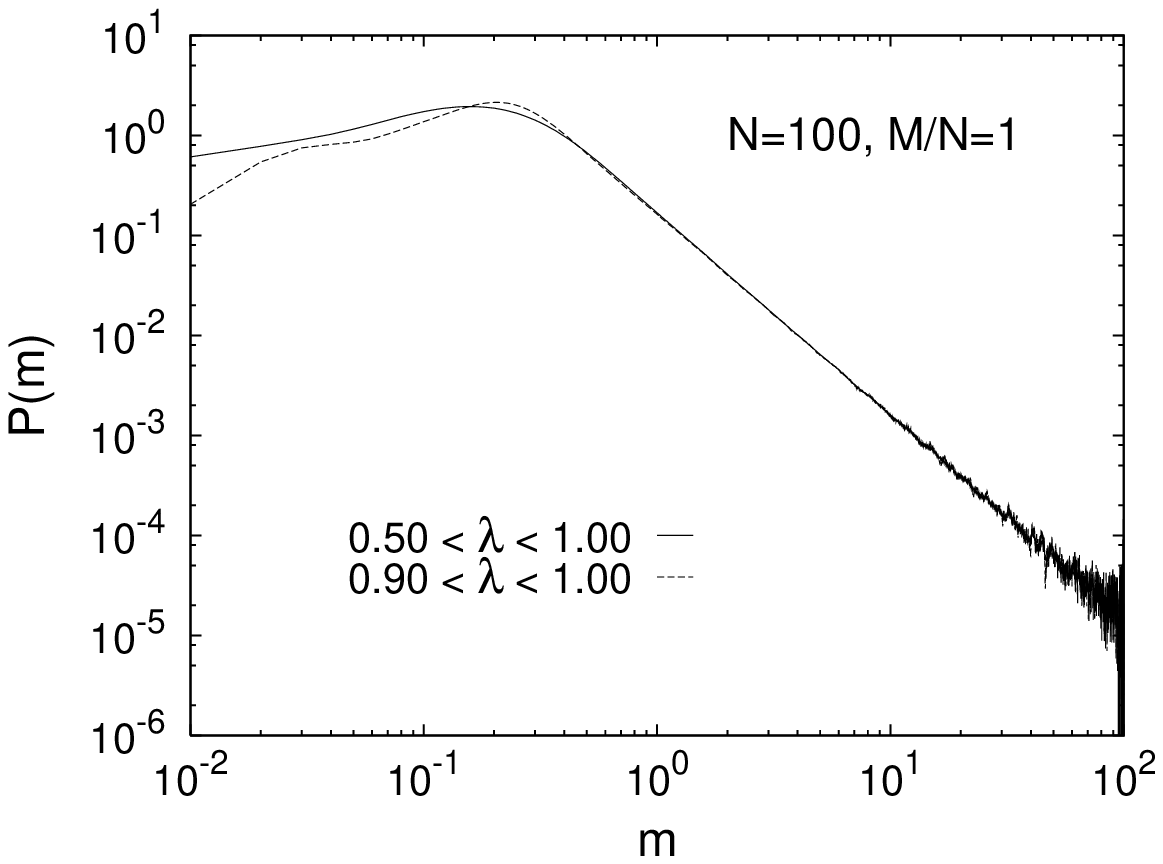}
}
\resizebox{0.99\columnwidth}{!}{
\includegraphics{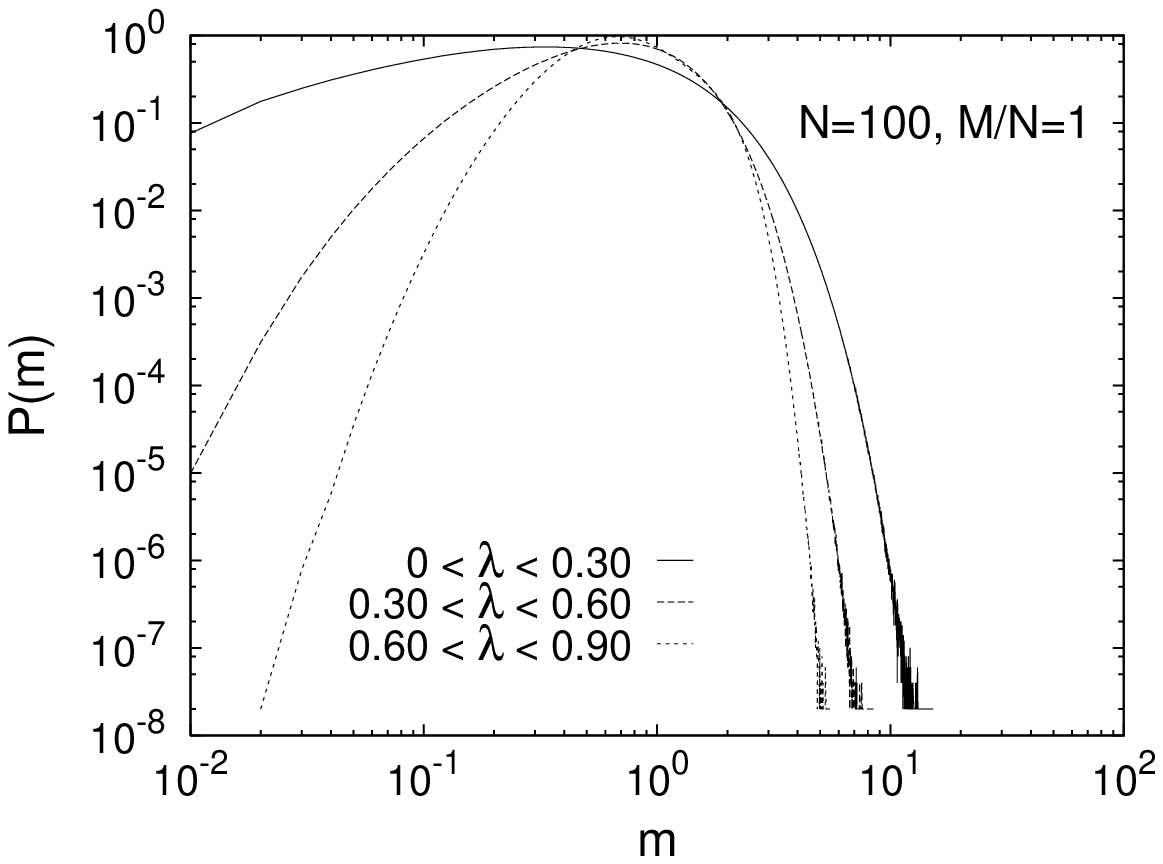}
}
\caption{Steady state money distribution in cases when the saving propensity
$\lambda$ is distributed uniformly within a range of values:
(a) $\lambda$ distribution extends upto $1$, money distribution shows
power law both for lower cut-offs $0.5$ and $0.9$;
(a) width of $\lambda$ distribution is $0.3$, money distribution shows
a power law in a narrow region only for $0.6 < \lambda < 0.9$.
The power law exponent is $\nu \simeq 1$ in all cases. 
All data shown here are for $N=100$, $M/N=1$.
}\label{lamrange}       
\end{figure}

We investigated on the range of distribution of the saving propensities in
a certain interval $a<\lambda_i<b$, where, $0<a<b<1$. For uniform distribution
within the range, we observe the
appearance of the same power law in the distribution but for a narrower
region. As may be seen from Fig.~\ref{lamrange}, 
as $a \rightarrow b$, the power-law
behavior is seen for values $a$ or $b$ approaching more and more towards unity:
For the same width of the interval $|b-a|$, one obtains power-law (with
the same value of $\nu$) when $b \rightarrow 1$. This indicates that 
for fixed $\lambda$, $\lambda=0$ correspond to a Gibbs distribution, 
and one observes a power law in $P(m)$ when $\lambda$ has got a non-zero
width of its distribution extending upto $\lambda = 1$. It must be emphasized
at this point that we are talking about the limit $\lambda \to 1$, since
any agent having $\lambda =1$ will result in condensation of money with that
particular agent. 
The role of the agents with high saving propensity ($\lambda \to 1$) is crucial: 
the power law behavior is truely
valid upto the asymptotic limit if $\lambda = 1$ is included. 
Indeed, had we assumed $\lambda_0=1$ in Eqn.~(\ref{lam0}),
the Pareto exponent $\nu$ immediately switches over to $\nu=1+\alpha$.
Of course, $\lambda_0 \ne 1$ in Eqn.~(\ref{lam0}) leads to the universality of the
Pareto distribution with $\nu = 1$ (irrespective of $\lambda_0$ and $\alpha$).
Obviously,
$P(m) \sim \int_0^1 P_{\lambda}(m)\rho(\lambda)d\lambda$ $\sim$ $m^{-2}$ for
$\rho(\lambda)$ given by Eqn.~(\ref{lam0}) and $P(m) \sim m^{-(2+\alpha)}$ if
$\lambda_0=1$ in Eqn.~(\ref{lam0}) (for large $m$ values).

Another numerical study~\cite{BMM} analysed the average money of the agent
with the maximum savings factor $\langle m(\lambda_{\rm max}) \rangle$.
This study concludes on the time evolution of the money of this agent,
and finds a scaling behavior
\begin{equation}
\label{manna-scaling}
\left[ \langle m(\lambda_{\rm max}) \rangle/N \right]
(1- \lambda_{\rm max})^{0.725}
\sim \mathcal{G} \left[ t (1- \lambda_{\rm max}) \right].
\end{equation}
This implies that the stationary state for the agent with the maximum
value of $\lambda$ is reached after a relaxation time
\begin{equation}
\label{manna-relax}
\tau \propto (1- \lambda_{\rm max})^{-1}.
\end{equation}
The average money $\langle m(\lambda_{\rm max}) \rangle$ of this agent
is also found to scale as
\begin {equation}
[\langle m(\lambda_{\rm max}) \rangle/N]N^{-0.15}
\sim {\cal F}[(1-\lambda_{\rm max})N^{1.5}].
\end {equation}
The scaling function
${\cal F}[x] \to x^{-\kappa}$ as $x \to 0$ with $\kappa \approx 0.76$.
This means $\langle m(\lambda_{\rm max}) \rangle N^{-1.15}$
$\sim (1-\lambda_{\rm max})^{-0.76}N^{-1.14}$
or $\langle m(\lambda_{\rm max}) \rangle \sim
(1-\lambda_{\rm max})^{-0.76}N^{0.01}$.
Since for a society of $N$ traders $(1-\lambda_{\rm max}) \sim 1/N$
this implies
\begin{equation}
\langle m(\lambda_{max}) \rangle \sim N^{0.77}.
\end{equation}

These model income distributions $P(m)$ compare very well with the wealth
distributions of various countries: Data suggests Gibbs like distribution
in the low-income range (more than 90\% of
the population) and Pareto-like in the high-income 
range \cite{Levy:1997,Dragulescu:2001,Aoyama:2003}
(less than 10\% of the population) of various countries. 
In fact, we compared one model simulation of the market with saving
propensity of the agents distributed following Eqn.~(\ref{lam0}), with 
$\lambda_0=0$ and $\delta=-0.7$~\cite{Chatterjee:2004}.
The qualitative resemblance of the model income distribution with the real
data for Japan and USA in recent years is quite intriguing. In fact, for
negative $\delta$ values in Eqn.~(\ref{lam0}), the density of traders with low
saving propensity is higher and since $\lambda=0$ ensemble yields
Gibbs-like income distribution Eqn.~(\ref{gibbs}), we see an initial Gibbs-like
distribution which crosses over to Pareto distribution Eqn.~(\ref{par}) with
$\nu=1.0$
for large $m$ values. The position of the crossover point depends on the
value of $\alpha$. It is important to note that any distribution
of $\lambda$ near $\lambda=1$, of finite width, eventually gives Pareto law
for large $m$ limit. The same kind of crossover behavior (from Gibbs to Pareto)
can also be reproduced in a model market of mixed agents where $\lambda=0$
for a finite fraction of population and $\lambda$ is distributed uniformly over
a finite range near $\lambda=1$ for the rest of the population.

\section{Analytical studies}\label{sec:anaystudy}
There have been a number of attempts to study the uniform savings
model (Model B, Sec.~\ref{subsec:fixedsaving}) 
analytically (see e.g.,~\cite{Das:2003}), 
but no closed form expression for the steady state distribution
$P(m)$ has yet been arrived at.
Kar Gupta~\cite{KarGupta:2006} 
investigated the nature of the transition matrices
from the equations Eqn.~(\ref{mi}) and Eqn.~(\ref{mj})
and concluded that the effect of introducing a saving
propensity leads to a nonsingular transition matrix,
and hence a time irreversible state.

We review now some of the investigations on the steady state distribution 
$P(m)$ of money resulting from the equations Eqn.~(\ref{mi}) 
and Eqn.~(\ref{mj}) representing the trading and money dynamics
(Model C, Sec.~\ref{subsec:mixedsaving}) in the distributed savings case.
The dynamics of money distribution is solved in two limiting cases. In one
case, the evolution of the mutual money difference among the agents
is investigated and one
looks for a self-consistent equation for its steady state distribution.
In the other case, a master equation for the money distribution
function is developed~\cite{Chatterjee:2005,Chatterjee:2005a}.

\subsection{Distribution of money difference}
Clearly in the process as considered 
(dynamics defined by Eqns.~(\ref{mi}) and (\ref{mj})), 
the total money $(m_i+m_j)$
of the pair of agents $i$ and $j$ remains
constant, while the difference $\Delta m_{ij}$ evolves as
\begin{eqnarray}
\label{dDelmijtt}
(\Delta m_{ij})_{t+1} &\equiv& (m_i-m_j)_{t+1} =
\left( \frac{\lambda_i+\lambda_j}{2} \right)(\Delta m_{ij})_t \nonumber\\
&+&
\left( \frac{\lambda_i-\lambda_j}{2} \right)(m_i+m_j)_t \nonumber\\
&& + (2 \epsilon_{ij} -1)[(1-\lambda_i)m_i(t)+(1-\lambda_j)m_j(t)].\nonumber\\
&&
\end{eqnarray}
\begin{figure}
\centering{
\resizebox*{9.0cm}{!}{\includegraphics{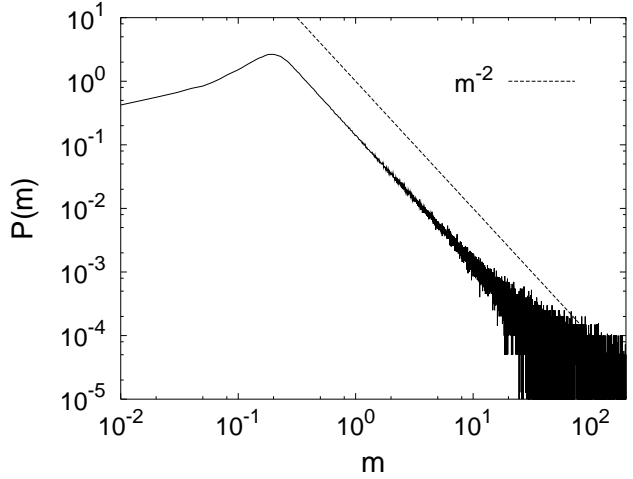}}
}
\caption{
Steady state money distribution $P(m)$ against $m$ in a numerical simulation
of a market with $N=200$, following equations Eqn.~(\ref{mi}) and Eqn.~(\ref{mj})
with 
$\epsilon_{ij}=1/2$.
The dotted line corresponds to $m^{-(1+\nu)}$; $\nu=1$.
Here, the average money per agent $M/N = 1$.
}\label{fig:latt}
\end{figure}
Numerically, as shown in Fig.~\ref{fig:distlambda}, 
we observe that the steady state money
distribution in the market becomes a power law, following such tradings
when the saving factor $\lambda_i$ of the agents remain constant over time
but varies from agent to agent widely. As shown in the numerical simulation
results for $P(m)$ in Fig.~\ref{fig:latt}, the law, as well as the exponent, 
remains unchanged even when $\epsilon_{ij}=1/2$ for every trading.
This can be justified by the earlier numerical observation
\cite{Chakraborti:2000,Chatterjee:2004} for fixed $\lambda$ market
($\lambda_i = \lambda$ for
all $i$) that in the steady state, criticality occurs as $\lambda \to 1$
where of course the dynamics becomes extremely slow. In other words,
after the steady state is realized, the third term in Eqn.~(\ref{dDelmijtt})
becomes unimportant for the critical behavior. For simplicity, 
we concentrate on this case, where the above evolution equation for
$\Delta m_{ij}$ can be written in a more simplified form as
\begin{equation}
\label{dDelmt}
(\Delta m_{ij})_{t+1} = \bar{\lambda}_{ij}(\Delta m_{ij})_t + \tilde{\lambda}_{ij}(m_i+m_j)_t,
\end{equation}
where $\bar{\lambda}_{ij}=\frac{1}{2}(\lambda_i+\lambda_j)$ and
$\tilde{\lambda}_{ij}=\frac{1}{2}(\lambda_i-\lambda_j)$. 
As such, $0 \le \bar{\lambda} < 1$ and
$-\frac{1}{2} < \tilde{\lambda} < \frac{1}{2}$.

The steady state probability distribution $D$ for the modulus
$\Delta = |\Delta m|$ of the mutual money difference between any two agents
in the market can be obtained from Eqn.~(\ref{dDelmt}) in the following way
provided $\Delta$ is very much larger than the average money per agent $=M/N$.
This is because, using Eqn.~(\ref{dDelmt}), large $\Delta$ can appear at
$t+1$, say, from `scattering' from any situation at $t$ for which the right
hand side of Eqn.~(\ref{dDelmt}) is large. The possibilities are (at $t$)
$m_i$ large (rare) and $m_j$ not large, where the right hand side of eqn.
Eqn.~(\ref{dDelmt}) becomes 
$\simeq (\bar{\lambda}_{ij} + \tilde{\lambda}_{ij})(\Delta_{ij})_t$;
or  $m_j$ large (rare) and $m_i$ not large (making the right hand side of eqn.
Eqn.~(\ref{dDelmt}) becomes 
$\simeq (\bar{\lambda}_{ij} - \tilde{\lambda}_{ij})(\Delta_{ij})_t$);
or when $m_i$ and $m_j$ are both large, which is a much rarer situation than
the first two and hence is negligible.
Consequently for large $\Delta$ the distribution $D$ satisfies
\begin{eqnarray}
\label{DDel}
D(\Delta)
&=& \int d \Delta^\prime \; D(\Delta^\prime) \; \nonumber\\
&& \times \langle
\delta (\Delta -(\bar{\lambda} + \tilde{\lambda}) \Delta^\prime) +
\delta (\Delta -(\bar{\lambda} - \tilde{\lambda}) \Delta^\prime)
\rangle \nonumber\\
&=&
2 \left <
\left( \frac{1}{\lambda} \right)
\; D
\left( \frac{\Delta}{\lambda} \right)
\right >,
\end{eqnarray}
where we have used the symmetry of the $\tilde{\lambda}$ distribution 
and the relation $\bar{\lambda}_{ij} + \tilde{\lambda}_{ij}=\lambda_i$, 
and have suppressed labels $i$, $j$.
Here $\langle \ldots \rangle$ denote average over $\lambda$ distribution
in the market, and $\delta$ denotes the $\delta$-function.
Taking now a uniform random distribution of the saving factor $\lambda$,
$\rho(\lambda) = 1$ for $0 \le \lambda < 1$, and assuming
$D(\Delta) \sim \Delta^{-(1+\nu)}$ for large $\Delta$, we get
\begin{equation}
\label{gammaex}
1=2 \int_0^1 d \lambda \; \lambda^\nu = 2 (1+\nu)^{-1},
\end{equation}
giving $\nu=1$.
No other value fits the above equation. This also
indicates that the money distribution $P(m)$ in the market also follows a
similar power law variation, $P(m) \sim m^{-(1+\nu)}$ and $\nu=1$.
Distribution of $\Delta$ from numerical simulations 
also agree with this result.

A detailed analysis of the master equation for the kinetic
exchange process and its solution for a special case
can be seen in Ref.~\cite{Chatterjee:2005,Chatterjee:2005a}.
For a pioneering study of the kinetic equations for
the two-body scattering process and a more general solution,
see Ref.~\cite{Repetowicz:2005,Richmond:2005}.

\subsection{A mean field explanation}

One can also derive the above results in a mean field limit, where
the money redistribution equations for the individual agents
participating in a trading process can be reduced to a stochastic map
in $m^2$~\cite{Bhattacharyya:2007}. 
The trick is to take the product of Eqn.~(\ref{mi}) and Eqn.~(\ref{mj})
and look for the time evolution of $m^2$:
\begin{eqnarray}
m_i(t+1) m_j(t+1) &=& \alpha_i (\epsilon_t,\lambda_i) m_i^2(t)
+ \alpha_j (\epsilon_t,\lambda_j) m_j^2(t) \nonumber\\
&& + \alpha_{ij} (\epsilon_t,\lambda_i, \lambda_j) m_i(t) m_j(t).
\label{eq:evolution-ij}
\end{eqnarray}
Since $\epsilon_{ij}$ in 
eqns.~(\ref{deltam}), (\ref{eps}) and (\ref{lrand}) 
keeps on changing (with time $t$) with the pairs of
scatterer ($i,j$), we use here $\epsilon_t$ to denote its explicit
time dependence.
We now introduce a mean-field-like approximation by replacing
each of the quadratic quantities $m_i^2$, $m_j^2$ and $m_i m_j$ by a mean
quantity $m^2$. Therefore Eqn.~(\ref{eq:evolution-ij}) is replaced by its
mean-field-like approximation
\begin{equation}
m^2(t+1) = \eta(t) m^2(t)
\label{eq:stochastic-map}
\end{equation}
where $\eta(t)$ is an algebraic function of $\lambda_i$,
$\lambda_j$ and $\epsilon_t$; it has been observed in numerical
simulations of the model that the value of $\epsilon_t$, whether it is
random or constant, has no effect on the steady state distribution
\cite{Chatterjee:2005} and the time dependence
of $\eta(t)$ results from the different values of $\lambda_i$ and
$\lambda_j$ encountered during the evolution of the market. Denoting
$\log (m^2)$ by $x$, Eqn.~(\ref{eq:stochastic-map}) can be written as
\begin{equation}
x(t+1) = x(t) + \delta(t),
\label{eq:random-walk}
\end{equation}
where $\delta(t) = \log \eta(t)$ is a random number that changes with
each time-step. The transformed map (Eqn.~(\ref{eq:random-walk})) depicts a
random walk and therefore the `displacements' $x$ in the time interval
$[0,t]$ follows the normal distribution
\begin{equation}
\mathcal{P}(x) \sim \exp \left ( -{x^2 \over t} \right ).
\label{eq:RW-distribution}
\end{equation}

\noindent Now
\begin{equation}
\mathcal{P} (x) \mathrm{d} x \equiv P(m) \mathrm{d} m^2
\label{eq:equivalence1}
\end{equation}

\noindent where $P(m)$ is the log-normal distribution of $m^2$:
\begin{equation}
P(m) \sim {1 \over m^2} \exp \left [ -{\left ( \log(m^2) \right )^2 \over t}
 \right ].
 \label{eq:log-normal1}
 \end{equation}
The normal distribution in Eqn.~(\ref{eq:RW-distribution}) spreads
with time (since its width is proportional to $\sqrt{t}$) and so does the
normal factor in Eqn.~(\ref{eq:log-normal1}) which eventually becomes a very
weak function of $m$ and may be assumed to be a constant as $t \to \infty$.
Consequently $P(m)$ assumes the form of a simple power law:
\begin{equation}
P(m) \sim {1 \over m^2} \ \mathrm{for} \ t \to \infty,
\label{eq:power-law1}
\end{equation}
that is clearly the Pareto law for the model.
Hence, the power law behavior obtained here agrees with the simulation
results. 

\subsection{Average money at any saving propensity and the distribution}

Several numerical studies 
investigated~\cite{Patriarca:EWD:2005,Patriarca:2006} 
the saving factor $\lambda$ and the average money held by an agent
whose savings factor is $\lambda$. This numerical study revealed
that the product of this average money and the unsaved fraction
remains constant, or in other words, the quantity
\begin{equation}
\langle m(\lambda) \rangle (1-\lambda) = c
\label{m1-lambda}
\end{equation} 
where $c$ is a constant.
This key result has been justified using a rigorous analysis by 
Mohanty~\cite{Mohanty:2006}.
We give below a simpler argument and proceed to derive the steady
state distribution $P(m)$ in its general form.

In a mean field approach, one can calculate~\cite{Mohanty:2006}
the distribution for the ensemble average of money
for the model with distributed savings.
It is assumed that the distribution of money of a single agent over
time is stationary, which means that the time averaged value of money
of any agent remains unchanged independent of the initial value of
money.
Taking the ensemble average of all terms on both sides of Eqn.~(\ref{mi}),
one can write
\begin{equation}
\label{pk-ensemble}
\langle m_i \rangle = \lambda_i \langle m_i \rangle +
\langle \epsilon \rangle
\left[
(1-\lambda_i) \langle m_i \rangle +
\langle \frac{1}{N} \sum_{j=1}^{N} (1-\lambda_j) m_j \rangle
\right].
\end{equation}
It is assumed that any agent on the average,
interacts with all others in the system.
The last term on the right is replaced by the average over the agents.
Writing
\begin{equation}
\overline {\langle (1-\lambda) m \rangle} \equiv
\left < \frac{1}{N} \sum_{j=1}^{N} (1-\lambda_j) m_j \right >
\end{equation}
and since $\epsilon$ is assumed to be distributed randomly and
uniformly in $[0,1]$, so that $\langle \epsilon \rangle = 1/2$,
Eqn.~(\ref{pk-ensemble}) reduces to
\[
(1-\lambda_i) \langle m_i \rangle = \overline {\langle (1-\lambda) m \rangle}.
\]
Since the right side is free of any agent index, it suggests that
this relation is true for any arbitrary agent, i.e.,
$\langle m_i \rangle (1-\lambda_i) =$ constant, where $\lambda_i$ is the
saving factor of the $i$th agent (as in Eqn.~(\ref{m1-lambda})) and
what follows is: 
\begin{equation}
\label{pk-ensemble2}
d \lambda \propto \frac{dm}{m^{2}}.
\end{equation}
An agent with a particular
saving propensity factor $\lambda$ therefore ends up with a
characteristic average wealth $m$ given by Eqn.~(\ref{m1-lambda}) 
such that one can in general relate the distributions of the two:
\begin{equation}
P(m) \ dm = \rho(\lambda) \ d\lambda.
\end{equation}
This, together with Eqn.~(\ref{m1-lambda}) and Eqn.~(\ref{pk-ensemble}) 
gives~\cite{Mohanty:2006}
\begin{equation}
P(m) = \rho(\lambda) \frac{d \lambda}{d m} 
	\propto \frac{\rho(1-\frac{c}{m})}{m^2},
\end{equation}
giving $P(m) \sim m^{-2}$ for large $m$
for uniform distribution of savings factor $\lambda$, i.e, $\nu=1$;
and $\nu = 1+ \delta$ for
$\rho(\lambda) = (1 - \lambda)^{\delta}$.
This study therefore explains the origin of the universal
($\nu=1$) as well as the non-universal ($\nu = 1+\delta$) 
Pareto exponent values in the distributed savings model, 
as discussed in Sec.~\ref{sec:ds:numerical}
and shown in Fig~\ref{fig:lamalpha} and Fig.~\ref{fig:1-al}.

\section{Other model studies}\label{sec:othmodel}
Sinha~\cite{Sinha:2003,Sinha:2005} considered an iterative map approach to
distribution of wealth in an economy, along with models that employed
yard-sale (YS) as well as theft and fraud (TF)~\cite{Hayes:2002} 
for asset exchange, yielding interesting results. 
A recent study~\cite{Saif:2007} also considers combinations of these
strategies, along with partial savings in a class of models.
Recent detailed studies~\cite{Yuqing:2007} of empirical data
and analysis of the distribution functions 
present a strong case in favor of gas-like models for economic exchanges.
Other studies calculated the holding time~\cite{Wang:2003} of money,
which indicated in turn the mobility of the money in a model under
a given dynamics. Another similar study~\cite{Wang:2005} calculated 
the velocity of money in a life-cycle model.
Studies of gas-like or particle-exchange models have already been
carried out on complex networks~\cite{Hu:2006,Hu:2007}.
Similar models study the effect of risk aversion and subsequent
emergence of Gibbs and power-law distributions in different
cases~\cite{Iglesias:2004}, while another study tunes the
rate of money transfer to obtain Boltzmann and Gibbs-like
money distributions~\cite{Ferrero:2004}. Similarly, one
can introduce asymmetry in favor of either of the traders
in a trade-investment framework and produce power law distributions
in wealth distributions~\cite{Scafetta:2004}.
Preferential spending behavior can also lead to similar
results~\cite{Ding:2003}.
Recently, Angle~\cite{Angle:EMBN:2007} has also proposed a
macro-model for the inequality process to explain the upward
surge of the Pareto tail in recent time for the US waged income data.
D\"{u}ring and Toscani~\cite{During:2007} recently formulated
hydrodynamic equations for such kinetic models of markets.

There are evidences of emerging income inequality arising as a consequence 
of resource flow in hierarchical organizations~\cite{Sinha:EMBN:2007},
and the resulting income distribution is power law distributed.

\subsection{A model with `annealed' savings}
In a real trading process, the concept of `saving factor' cannot be 
attributed to a quantity that is invariant with time. A saving factor 
always changes with time or trading.
In some of the earlier works~\cite{Chatterjee:2004}, 
we reported the case of annealed savings,
where the savings factor $\lambda_i$ changes with time in the
interval $[0,1)$, but does not give rise to a power law in
$P(m)$~\cite{Chatterjee:2004}. 
But, there are some special cases
of annealed saving can give rise to a power law distribution of $P(m)$.
\begin{figure}
\resizebox{0.99\columnwidth}{!}{
\includegraphics{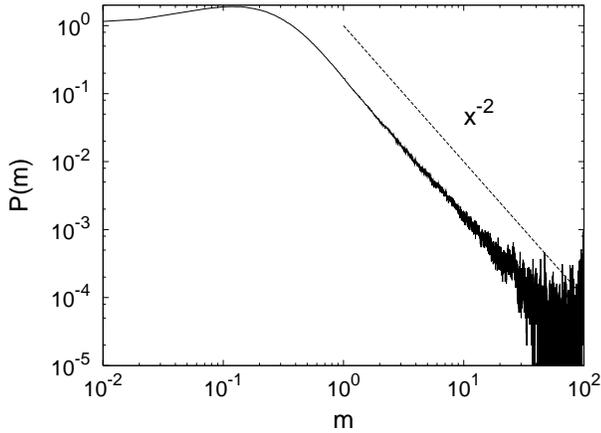}
}
\caption{
Distribution $P(m)$ of money $m$ in case of annealed savings $\lambda$
varying randomly in $[\mu,1)$. Here, $\zeta(\mu)$ has a uniform distribution.
The distribution produces a power law tail with Pareto exponent $\nu=1$.
The simulation has been done for a system of $N=100$ agents, with 
average money per agent $M/N=1$.
$P(m)$ is the steady state distribution after $4 \times 10^4$ Monte Carlo
steps, and the data is averaged over an ensemble of $10^5$.
}
\label{fig:ann:1-lambda}
\end{figure}

We proposed~\cite{ecoanneal} 
a slightly different model of an annealed saving case.
Let us associate a parameter $\mu_i$ ($0 < \mu_i < 1$) with each agent $i$
such that the savings factor $\lambda_i$ randomly assumes a value in the
interval $[\mu_i,1)$ at each time or trading.
The trading rules are of course unaltered and governed by 
Eqns.~(\ref{mi}) and (\ref{mj}).
Now, considering a suitable distribution $\zeta(\mu)$ of $\mu$
over the agents, one can produce money distributions with power-law
tail. The only condition that needs to be satisfied is that $\zeta(\mu)$
should be non-vanishing as $\mu \to 1$.
Figure~\ref{fig:ann:1-lambda} shows the case when $\zeta(\mu)=1$.
Numerical simulations suggest that the behavior of the wealth
distribution is similar to the quenched savings case. In other
words, only if $\zeta(\mu) \propto |1-\mu|^\delta$, it is reflected
in the Pareto exponent as $\nu=1+\delta$~\cite{ecoanneal}.
$\mu_i$ is interpreted as the lower bound of the saving distribution
of the $i$-th agent. Thus, while agents are allowed to randomly save any
fraction of their money, the bound ensures that there is always a
non-vanishing fraction of the population that assumes high saving fraction.

\subsection{A model with a non-consumable commodity}
\label{subsec:nonconsum}
Money is certainly not the only quantity that circulates in a trading market.
Exchange of goods is the main entity for transactions. Different economic
conditions give rise to the fluctuation of price of these commodities
and this plays an important role in the behavior of the market as a whole.
The determination of `price' is a complex phenomena and is decided
by the dynamics of supply and demand of the particular commodity.

In the trading markets discussed in previous two chapters, 
modifications due to exchange of a consumable
commodity hardy affects the distribution, as the commodity once bought
or sold need not be accounted for.
Consumable commodities effectively have no `price', as due to their short
lifetime to contribute to the total wealth of an individual.
It is interesting however, to study the role of non-consumable commodities
in such market models.
\begin{figure}
\resizebox{0.99\columnwidth}{!}{
\includegraphics{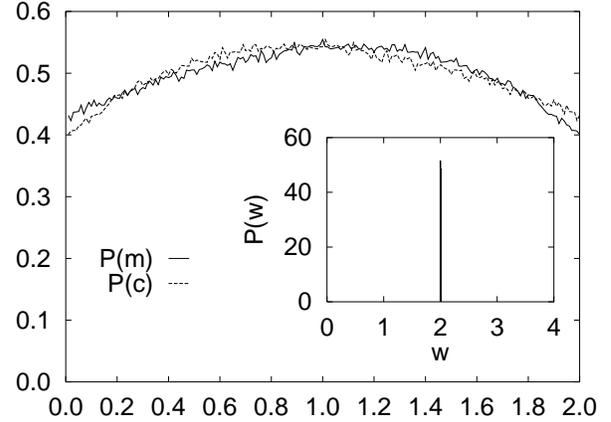}
}
\caption{
Steady state distribution $P(m)$ of money $m$ in a market with no savings
(saving factor $\lambda=0$) for no price fluctuations i.e, $\theta=0$.
The graphs show simulation results for a system of $N=100$ agents,
$M/N=1$, $C/N=1$; $m_i=c_i=1$ at $t=0$ for all agents $i$.
The inset shows the distribution $P(w)$ of total wealth $w=m+c$.
As $p=1$, for $\theta=0$, although $m$ and $c$ can change with tradings
within the limit $(0-2)$ the sum is always maintained at $2$.
}
\label{fig:commogif.flam.del0}
\end{figure}

For sake of simplicity, we consider a  simplified version of a market with 
a single non-consumable commodity~\cite{Chatterjee:2006}. 
As before, we consider a fixed number of traders or agents $N$
who trade in a market with total money $\sum_i m_i(t)=M$ and total
commodity $\sum_i c_i(t)=C$, 
$m_i(t)$ and $c_i(t)$ being the money and commodity respectively
of the $i$-th agent at time $t$ and are both non-negative.
Needless to mention, both $m_i(t)$ and $c_i(t)$ change with time 
or trading $t$.
The market, as in previous cases, is closed, i.e., $N$, $M$ and $C$
are constants.
The wealth $w_i$ of an individual $i$ in that case is, the sum of the money
and commodity it possesses, i.e., $w_i=m_i + p_0 c_i$; where
$p_0$ is the ``global'' price.
In course of trading, total money and total commodity are locally conserved,
and this automatically conserves the total wealth.
In such a market, one can define a global average price parameter
$p_0=M/C$, which is set to unity in this case, giving $w_i = m_i + c_i$.
It may be noted at this point that in order to avoid the complication of 
restricting the commodity-money exchange and their reversal between 
the same agents, the Fisher velocity of money circulation 
(see e.g., Ref.~\cite{Wang:ESTP}) is renormalised to unity here.
In order to accommodate the lack
of proper information and the ability of the agents to bargain etc., we will
allow fluctuations $\theta$ in the price of the commodities
at any trading (time): $p(t)= p_0 \pm \theta = 1 \pm \theta$.
We find, the nature of steady state to be unchanged and independent of
$\theta$, once it becomes non-vanishing.

In general, the dynamics of money in this market looks the same as
Eqns.~(\ref{mdelm}), (\ref{deltam}), (\ref{fmi}), (\ref{fmj}), (\ref{eps}) or
(\ref{mi}), (\ref{mj}), (\ref{lrand}) depending on whether
$\lambda_i=0$ for all, $\lambda_i \ne 0$ but uniform for all $i$ or
$\lambda_i \ne \lambda_j$ respectively.
However, all $\Delta m$ are not allowed here; only those, for which
$\Delta m_i \equiv m_i(t+1)-m_i(t)$ or $\Delta m_j$ are allowed by the
corresponding changes $\Delta c_i$ or $\Delta c_j$
in their respective commodities ($\Delta m > 0, \Delta c > 0$)~\cite{Chatterjee:2006}:
\begin{equation}
\label{commo-eqn1}
c_i(t+1)=c_i(t)+ \frac{m_i(t+1)-m_i(t)}{p(t)}
\end{equation}
\begin{equation}
\label{commo-eqn2}
c_j(t+1)=c_j(t)- \frac{m_j(t+1)-m_j(t)}{p(t)}
\end{equation}
where $p(t)$ is the local-time `price' parameter, a stochastic variable:
\begin{equation}
\label{pricedelta}
p(t)=
\left\{\begin{array}{c}
1 + \theta {\rm \;\; with \; probability \;0.5}\\
1 - \theta {\rm \;\; with \; probability \;0.5}
\end{array}\right..
\end{equation}
The role of the stochasticity in $p(t)$ is to imitate the effect of
bargaining in a trading process. $\theta$ parametrizes the amount of
stochasticity. The role of $\theta$ is significant in the sense that it
determines the (relaxation) time the whole system takes to reach a dynamically
equilibrium state; the system reaches equilibrium sooner for larger
$\theta$, while its magnitude does not affect the steady state distribution.
It may be noted that, in course of trading process, certain exchanges
are not allowed (e.g., in cases when a particular pair of traders do not
have enough commodity to exchange in favor of an agreed exchange of money).
We then skip these steps and choose a new pair of agents for trading.

In an ideal gas market without savings, money is
exponentially distributed in presence of any finite value of $\theta$.
Again, commodity has a small initial peak before decaying exponentially.
However, the total wealth $w=m+c$ has a form of a Gamma distribution.

For $\theta=0$, however, wealth of each agent remains invariant with time
as only the proportion of money and commodity interchange within themselves,
as the `price' factor remains constant.
This of course happens irrespective of the savings factor being zero,
uniform or distributed. For $\theta=0$, the steady state distribution
of money or commodity can take non-trivial forms:
see Fig.~\ref{fig:commogif.flam.del0}, but strictly a $\delta$-function
for total wealth, or at the value of wealth one starts with
(see inset of Fig.~\ref{fig:commogif.flam.del0} for the case $m_i=c_i=1$
for all $i$)~\cite{Chatterjee:2006}.
\begin{figure}
\resizebox{0.99\columnwidth}{!}{
\includegraphics{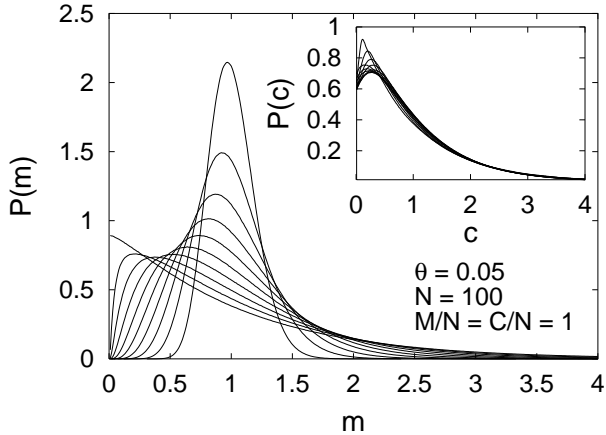}
}
\caption{
Steady state distribution $P(m)$ of money $m$ in the
uniform savings commodity market
for different values of saving factor $\lambda$
($0,0.1,0.2,0.3,0.4,0.5,0.6,0.7,0.8,0.9$ from left to right near the origin)
for $\theta=0.05$.
The inset shows the distribution $P(c)$ of commodity $c$ in the uniform
savings commodity market for different values of saving factor $\lambda$.
The graphs show simulation results for a system of
$N=100$ agents, $M/N=1$, $C/N=1$.
}
\label{fig:commogif.flam.mc}
\end{figure}
\begin{figure}
\resizebox{0.99\columnwidth}{!}{
\includegraphics{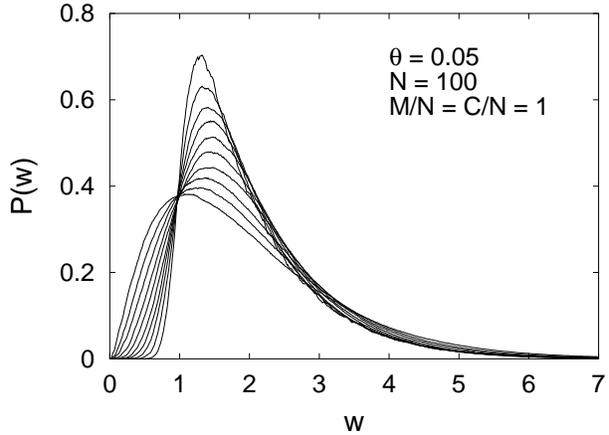}
}
\caption{
Steady state distribution $P(w)$ of total wealth $w=m+c$
in the uniform savings commodity
market for different values of saving factor $\lambda$
($0,0.1,0.2,0.3,0.4,0.5,0.6,0.7,0.8,0.9$ from left to right)
for $\theta=0.05$.  The graphs show simulation results for a system of
$N=100$ agents, $M/N=1$, $C/N=1$.
}
\label{fig:commogif.flam.w}
\end{figure}

As mentioned already for $\theta \ne 0$, the steady state results
are not dependent on the value of $\theta$, the relaxation time
of course decreases with increasing $\theta$.
In such a market with uniform savings, money distribution $P(m)$ has a form
similar to a set (for $\lambda \ne 0$)
of Gamma functions (see Fig.~\ref{fig:commogif.flam.mc}):
a set of curves with a most-probable value shifting from $0$ to $1$
as saving factor $\lambda$ changes from  $0$ to $1$
(as in the case without commodity).
The commodity distribution $P(c)$ has an initial peak and an exponential
fall-off, without much systematics with varying
$\lambda$ (see inset of Fig.~\ref{fig:commogif.flam.mc}).
The distribution $P(w)$ of total wealth $w=m+c$ behaves much like $P(m)$
(see Fig.~\ref{fig:commogif.flam.w}).
It is to be noted that since there is no precise correspondence with
commodity and money for $\theta \ne 0$
(unlike when $\theta=0$, when the sum is fixed),
$P(w)$ cannot be derived directly from $P(m)$ and $P(c)$.
However, there are further interesting
features. Although they form a class of Gamma distributions, the set of
curves for different values of saving factor $\lambda$ seem to intersect
at a common point, near $w=1$. All the reported data are for a system
of $N=100$ agents, with $M/N=1$ and $C/N=1$ and for a case where the noise
$\theta$ equals $0.5$~\cite{Chatterjee:2006}.

\begin{figure}
\resizebox{0.99\columnwidth}{!}{
\includegraphics{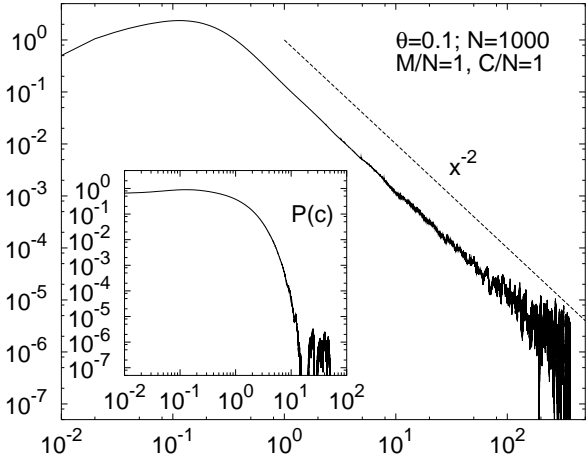}
}
\caption{
Steady state distribution $P(m)$ of money $m$
in the commodity market with distributed savings $0 \le \lambda < 1$.
$P(m)$ has a power-law tail with Pareto exponent $\nu=1 \pm 0.02$
(a power law function $x^{-2}$ is given for comparison).
The inset shows the distribution $P(c)$ of commodity $c$
in the same commodity market.
The graphs show simulation results for a system of
$N=1000$ agents, $M/N=1$, $C/N=1$.
}
\label{fig:commogif.dlam.mc}
\end{figure}

\begin{figure}
\resizebox{0.99\columnwidth}{!}{
\includegraphics{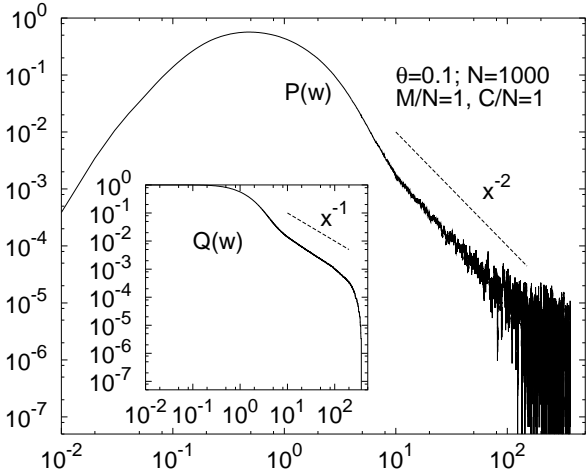}
}
\caption{
Steady state distribution $P(w)$ of total wealth $w=m+c$
in the commodity market with distributed savings $0 \le \lambda < 1$.
$P(w)$ has a power-law tail with Pareto exponent $\nu=1 \pm 0.05$
(a power law function $x^{-2}$ is given for comparison).
The inset shows the cumulative distribution $Q(w) \equiv \int_w^\infty P(w) {\rm d}w$.
The graphs show simulation results for a system of
$N=1000$ agents, $M/N=1$, $C/N=1$.
}
\label{fig:commogif.dlam.w}
\end{figure}

For $\lambda$ distributed uniformly within the interval $0 \le \lambda < 1$,
the tails of both money and wealth distributions $P(m)$ and $P(w)$ have
Pareto law behavior with a fitting exponent value
$\nu=1 \pm 0.02$ and $\nu=1 \pm 0.05$ respectively
(see Fig.~\ref{fig:commogif.dlam.mc} and Fig.~\ref{fig:commogif.dlam.w}
respectively),
whereas the commodity distribution is still exponentially decaying
(see inset of Fig.~\ref{fig:commogif.dlam.mc})~\cite{Chatterjee:2006}.


A major limitation of these money-only exchange models
considered earlier~\cite{EWD05,Ispolatov:1998,marjit,Dragulescu:2000,Chakraborti:2000,Hayes:2002,Chatterjee:2004,Chatterjee:2003,Chakrabarti:2004,Slanina:2004,ESTP:KG,EWD:CC,CurrSci,Angle:1986,Lux:EWD:2005,Angle:2006,Patriarca:2004,Repetowicz:2005,Richmond:2005,Mohanty:2006,Chatterjee:2005,ecoanneal}
is that they do not make any explicit reference to the commodities 
exchanged with the money and to the constraints they impose on the 
exchange process.
Also, the wealth is not just the money is possession (unless the commodity
exchanged with the money is strictly consumable).
Here, we have studied the effect
of a single non-consumable commodity on the money (and also wealth)
distributions in the steady state, and allowing for local (in time)
price fluctuation.
Allowing for price fluctuation is very crucial for the model --
it allows for the stochastic dynamics to play its proper role
in the market.
However, this model is quite different from that considered
recently in Ref.~\cite{Ausloos:2006}, where $p_0$ is strictly unity
and the stochasticity enters from other exogenous factors.
In the sense that we also consider two exchangeable variables in the market,
our model has some similarity with that in Ref~\cite{Silver:2002}.
However, Silver et al~\cite{Silver:2002}
consider only random exchanges between agents (keeping the total conserved)
while we consider random exchanges and also allowing for price fluctuations
and savings. As such they only obtain
the Gamma distribution in wealth, while our model produce 
both Gamma and Pareto distributions.
In spite of many significant effects,
the general feature of Gamma-like form of the money (and wealth)
distributions (for uniform $\lambda$) and the power law tails
for both money and wealth (for distributed $\lambda$)
with identical exponents, are seen to
remain unchanged. The precise studies (theories) for the money-only
exchange models are therefore extremely useful and relevant.
\section{Discussions}
\label{sec:disc}
Empirical data for income and wealth distribution in many countries
are now available, and they reflect a particular robust 
pattern (see Fig.~\ref{fig:realdataset}).
The bulk (about 90\%) of the distribution resemble the century-old 
Gibbs distribution of energy for an ideal gas, while there are evidences 
of considerable deviation in the low income as well as high income ranges.
The high income range data (for 5-10\% of the population in any country) 
fits to a power law tail, known after Pareto, and the value of the 
(power law) exponent ranges between 1-3 and depends on the individual
make-up of the economy of the society or country.
There are also some reports of two distinct power law tails
of such distributions (see e.g.~\cite{Richmond:EMBN:2007}).

The analogy with a gas like many-body system has led
to the formulation of the models of markets.
The random scattering-like dynamics of money (and wealth) in a closed 
trading market, in analogy with energy conserved exchange models,
reveals interesting features.
The minimum modification required over such ideal gas-like kinetic
exchange models seem to be the consideration of 
saving propensity of the traders. 
Self-organisation is a key emerging feature of these kinetic exchange
models when saving factors are introduced.
In the model with uniform savings (see Sec.~\ref{subsec:fixedsaving}), 
the Gamma-like distribution of wealth shows stable 
most-probable or peaked distribution
with a most-probable value indicative of an economic scale dependent
on the saving propensity or factor $\lambda$. Empirical
observations in homogeneous groups of individuals as in waged income
of factory labourers in UK and USA~\cite{Willis:2004}
and data from population survey in USA among students of different school
and colleges produce similar distributions~\cite{Angle:2006}.
This is a relatively simpler case where a homogeneous population
(say, characterised by a unique value of $\lambda$) could be identified.

In the model with distributed savings (see Sec.~\ref{subsec:mixedsaving}), 
the saving propensity is assumed to have a randomness and varies from 
agent to agent. 
One finds the emergence of a
power law tail in money (and wealth) in cases where the saving
factor is a quenched variable (does not change with tradings or time $t$) 
within different agents or traders.
Several variants have been investigated for
the basic model, including an `annealed' version, some of which produce
the Pareto-like power law (Eqn.~(\ref{par})). 
The money exchange equations can be cast into
a master equation, and the solution to the steady state money distribution
giving the Pareto law with $\nu=1$ have been derived using
several approaches (see Sec.~\ref{sec:anaystudy}).
The results of the mean field theory agree with the simulations.
We have mostly used the terms `money' and `wealth' interchangably,
treating the models in terms of only one quantity, namely `money' that
is exchanged. Ofcourse, wealth does not comprise of (paper)
money only, and there have been studies distinguishing these two.
We review one such model study in Sec.~\ref{subsec:nonconsum} where, 
in addition to money, a single non-consumable commodity, 
having local price fluctuations, was introduced. The steady state
money and wealth (comprising of money and price weighted commodity) 
distributions were then investigated in the same market.
Interestingly, the scaling behavior for high range of the money as well as the
wealth are found to be similar (see Sec.~\ref{subsec:nonconsum}),
with identical Pareto exponent value for the distributed savings.

Study of such simple models here give some insight into the possible
emergence of self organizations in such markets, evolution of the steady 
state distribution, emergence of Gamma-like distribution for the bulk
and of the power law tail, as in the empirically observed 
distributions (Fig.~\ref{fig:realdataset}). 
A study of these models in terms of quantities that parametrise
the circulation of money~\cite{Wang:ESTP} suggests that
the model with distributed savings perform better.
These studies bring some new insight into the some essential
economic issues, including economic mobility.

These model studies also indicate
the appearance of self-organization, and the self-orgaized 
criticality~\cite{Bak:1997} in particular, in the simplest
model so far; namely in the kinetic gas models, when the
effect of random saving propensities \cite{Samuelson:1980} is incorporated.
Our observations indicate that the Gibbs and the (self-organized critical) 
Pareto distributions fall in the same category and can appear
naturally in the century-old and well-established kinetic theory of gas
\cite{Landau:1968,ESOM:2006}:
Gibbs distribution for no saving and Pareto distribution for
agents with quenched random saving propensity. 
To some degree of approximation therefore, these studies indicate that
the society or market behaves like an ideal gas, and the exchange of money
and wealth looks similar as in the above models at a coarse-grained level.
Statistical physics allows us to model and analyse such systems
in analogy to a variety of many body systems studied traditionally
within the framework of physics; see e.g., Yakovenko \cite{Yako:Encyclo} 
for an alternative account on these developments.

These models have additional prospective future applications in 
other spheres of social as well as
physical sciences. In social sciences, the knowledge of the mechanism
by which such distributions of wealth emerge out of collective exchanges
may find application in policy making and taxation~\cite{Hogan}.
In physical sciences, the corresponding particle exchange model
can find important application in designing desired energy 
spectrum for different types of chemical reactions~\cite{Scafetta:2007}.


\begin{acknowledgement}
The authors are grateful to
P.~Bhattacharyya, A.~Chakraborti, S.~S.~Manna, S.~Marjit, S.~Pradhan,
S.~Sinha and
R.~B.~Stinchcombe for collaborations at various stages of the study.
Useful discussions with
J.~Angle, A.~S.~Chakrabarti, A.~Das, Y.~Fujiwara, M.~Gallegati, A.~Kar~Gupta,
T.~Lux, M.~Marsili, P.~K.~Mohanty, L.~Pietronero,
P.~Richmond, W.~Souma, D.~Stauffer, V.~M.~Yakovenko and S.~Yarlagadda
are also acknowledged.
\end{acknowledgement}

\end{document}